\documentclass[preprint]{emulateapj}
\usepackage{lscape} 

\slugcomment{To be Published in the Astrophysical Journal}
\shortauthors{Zaritsky, Gonzalez, \& Zabludoff}
\shorttitle{Fundamental Manifold of Spheroids}

\received{2005 May 29}
\begin{document}
\title{
The Fundamental Manifold of Spheroids
}

\author{Dennis Zaritsky\altaffilmark{1}, Anthony H. Gonzalez\altaffilmark{2,3}, and Ann I. Zabludoff\altaffilmark{1}}
\altaffiltext{1}{Steward Observatory, University of Arizona, 933 North Cherry Avenue, Tucson, AZ 85721}
\altaffiltext{2}{Department of Astronomy, University of Florida, Gainesville, FL 32611}
\altaffiltext{3}{NSF Astronomy and Astrophysics Postdoctoral Fellow}

\begin{abstract}    
We present a unifying empirical description of the structural and kinematic properties of 
all spheroids embedded in dark matter halos.
We find that the stellar spheroidal components of galaxy clusters,
which we call cluster spheroids (CSphs) and which are typically one hundred
times the size of normal elliptical galaxies,  lie on a ``fundamental plane" 
as tight as that defined by ellipticals (rms in effective radius of $\sim$ 0.07), but that
has a different slope. The slope, as measured by the
coefficient of the $\log \sigma$ term, declines significantly and 
systematically between the fundamental planes of ellipticals, 
brightest cluster galaxies (BCGs), and CSphs.
We attribute this decline primarily to a 
continuous change in $M_e/L_e$, the mass-to-light ratio within the effective radius $r_e$,
with spheroid scale. The magnitude of 
the slope change requires that it arises principally from differences in 
the relative distributions of 
luminous and dark matter, rather than from stellar population differences such as
in age and metallicity. By expressing the
$M_e/L_e$ term as a function of $\sigma$ in the simple derivation of the fundamental plane
and requiring
the behavior of that term to mimic the observed nonlinear 
relationship between $\log M_e/L_e$ and $\log \sigma$, we
simultaneously fit a 2-D manifold to the measured properties of dwarf ellipticals,
ellipticals, BCGs, and CSphs. The combined data have an rms scatter in $\log r_e$
of 0.114 (0.099 for the combination of Es, BCGs, and CSphs), which is
modestly larger than each fundamental plane has alone, but which includes the scatter
introduced by merging different studies done in different filters by different investigators. 
This ``fundamental manifold'' fits the structural and kinematic properties
of spheroids that span a factor of 100 in $\sigma$ and 1000 in
$r_e$.  While our mathematical
form is neither unique nor derived from physical principles, the tightness of the fit leaves little room for 
improvement by other unification schemes over the range of observed spheroids.

\end{abstract}


\section{Introduction}

Given the differences in how, where, and when galaxies form,
there are some relationships among galaxy parameters that appear impossibly
tight. The importance of such relationships is that they must evince 
deeper and yet unknown laws of galaxy formation. The principal example
of such a relationship is that relating scale $r_e$, surface brightness $I_e$, and velocity dispersion $\sigma$
for elliptical galaxies \citep{d87,dd87}. In the 3-space defined by these parameters,
elliptical galaxies lie on a plane with modest scatter \citep[$\sim$ 0.1 dex;][]{jorg,bernardi}.
Although the origin of this remarkable result is not 
fully understood, this relationship is widely known
as the ``Fundamental Plane" (hereafter, FP). 

There have been previous efforts to place all dynamically hot systems onto
the FP, or related relationships \citep[e.g.,][]{kormendy,burstein}.
Although 
systems from globular clusters \citep{burstein} to galaxy clusters \citep{schaeffer} do obey general
relationships similar to the elliptical FP, 
the slopes and intercepts are often significantly different, 
suggesting differences in the role of 
dissipation, the distribution of angular momentum, and the relative distributions
of light and dark matter.
Although we have yet to untangle the clues provided by these relationships, probing
the extremes of the family of FPs --- where spheroids must arise from qualitatively
different formation processes --- may provide a breakthrough.
For example, while dissipation is clearly important in boosting the central
phase space density of normal Es \citep{lauer, carlberg}, the hierarchical
accretion that produces the largest spheroids, the intracluster light component of
galaxy clusters, is mostly dissipationless \citep{murante, willman, sl}. Differences
may then arise in how spheroids across these scales populate $(r_e, I_e, \sigma)$-space.

Originally identified by \cite{mathews}, the light in excess of the standard
$r^{1/4}$ surface brightness profile observed at large radii in some brightest
cluster galaxies (BCGs) 
is generally referred to as the ``cD envelope", although specialists have 
a wider range of terminology \citep{oemler73,oemler76,
s86, s87, s88}. More recent work \citep{uson90,uson91,sk,g2000} questioned 
the existence of such a component and suggested that difficulties in background subtraction
at such low surface brightness levels might have skewed measurements. 
However, the latest efforts \citep{f02, f04, GZZ, zibetti} have all found excess luminosity 
above the r$^{1/4}$ profile of BCGs, 
although at a surface brightness well below that associated
with classical ``cD envelopes". 
This luminosity is attributed to intracluster stars (hereafter, ICS),
a stellar population that is gravitationally bound to the cluster but unbound
to any individual cluster galaxy. The ICS component is well described by an $r^{1/4}$ surface
brightness profile and satisfies a relationship between  $r_e$
and $I_e$ that is similar to that of normal
elliptical galaxies \citep[][hereafter Paper I]{GZZ}.  As such, the stellar component associated with
the cluster as a whole appears to be an extremely large version of normal spheroids, so we designate
it as the cluster spheroid (CSph).

To examine the place of CSphs among
more traditional 
spheroids requires a measurement of the CSph velocity dispersion in clusters
where we have already observed the ICS (Paper I).
Although the ideal experiment would involve measuring the 
velocity dispersion of the ICS directly, that task is quite difficult because of its low
surface brightness \citep[see][]{kelson} and is
not possible for our large sample of clusters without a tremendous investment of
large telescope time. Therefore, we adopt the cluster galaxy velocity dispersion as
the CSph dispersion and discuss possible shortcomings of this approach.
We review briefly
the status of the ICS measurements and present our cluster kinematic measurements
in \S \ref{sec:data}. In \S3 we discuss the connection between the BCG and ICS, the
FP populated by the CSphs, and a new relationship between the FPs of ellipticals,
BCGs, and CSphs.  In 
\S5 we summarize our findings.

\label{sec:intro}

\section{Data}

\label{sec:data}

In Paper I we present observations of 24 clusters and rich groups that span a range of velocity dispersions and Bautz-Morgan types \citep{bm}. The sample consists of nearby systems ($0.03 < z < 0.13$) that contain a dominant BCG with a major axis position angle that lies within 45$^\circ$ of the east-west axis (the drift scan direction). We present details of these unique data and 
the reduction procedure in Paper I.  Here we discuss the 23 of those clusters for which we 
obtained kinematic data (Abell 2376 is the one cluster from Paper I that is omitted here).
All calculated quantities in this paper are for
a concordance cosmology model:  $\Omega_m = 0.3$, $\Omega_\Lambda = 0.7$, and
$H_0 = 70$ km sec$^{-1}$ Mpc$^{-1}$.

We obtained spectra of galaxies in the fields of the 12 clusters from Paper I with
insufficient kinematic data in the literature.
We observed those galaxies with the Hydra spectrograph at the CTIO Blanco 4m telescope on the
nights of 28 July 2003 to 31 July 2003 using
the KPGL1 grating, which has 632 lines mm$^{-1}$ and provides a dispersion
of 0.59 \AA\  pixel$^{-1}$. The measured spectra span 3700 to 6084 \AA. In combination
with the bare fibers (no slit plate was used), the resulting spectral resolution is 4.1 \AA.
Targets were selected from our optical imaging (described in Paper I) to satisfy 
the simple criteria that they be fainter than the BCG and lie
within a projected distance of 1.5 Mpc from the BCG. 
Priority for fiber assignments was based
on galaxy magnitude. 
We obtained flat fields and wavelength calibrations through the fibers
at each target position and typically observed each cluster field 
for 1800 sec,  three to four times.
Total exposure times vary due to variable weather conditions, but the above numbers
are representative. 

\begin{deluxetable*}{lccccrrrrrr}
\tabletypesize{\tiny}
\tablecaption{Measured Cluster Parameters}
\tablehead{
\colhead{Cluster}& 
\colhead{$M_{ICS}$}& 
\colhead{$r_{e,ICS}$}&
\colhead{$M_{BCG}$}& 
\colhead{$r_{e,BCG}$}&
\colhead{$M_{e,GAL}$}&
\colhead{Source}&
\colhead{No.} &
\colhead{Range} &
\colhead{$\bar v$}&
\colhead{$\sigma$}\\
& 
& 
\colhead{[kpc]}&
& 
\colhead{[kpc]}&
&
&
&
&
\colhead{[km s$^{-1}$]} &
\\
}
\startdata
A0122  &$-25.56$&$107.9$&$-22.97$&$ 4.7$&$-24.97$&H&28&[32410,35611]&$34020\pm130$&$680^{+110}_{-90}$\\
&$\pm 0.01^{0.06}_{-0.06}$&$\pm 2.9^{9.3}_{-7.4}$&$\pm 0.03^{0.05}_{-0.05}$&$\pm 0.1^{0.2}_{-0.2}$\\
A1651  &$-24.95$&$477.2$&$-25.21$&$48.2$&   $-25.76$&N&28&[23338,27324]&$25400\pm190$&$990^{+110}_{-100}$\\
&$\pm 0.07^{0.48}_{-0.53}$&$\pm53.1^{239.1}_{-130.5} $&$\pm 0.01^{0.00}_{-0.02}$&$\pm 0.5^{0.6}_{-0.2}$\\
A2400  &$-24.61$&$184.6$&$-24.19$&$20.4$&   $-24.37$&H&50&[25147,27660]&$26380\pm90$&$650^{+60}_{-60}$\\
&$\pm 0.04^{0.19}_{-0.44}$&$\pm44.9^{169.7}_{-75.2}$&$\pm 0.08^{0.17}_{-0.11}$&$\pm 1.0^{0.8}_{-1.4}$\\
A2401  &$-24.48$&$ 27.6$&$-22.30$&$ 1.8$&   $-20.13$&N&24&[16123,18207]&$17120\pm100$&$460^{+100}_{-80}$\\
&$\pm 0.01^{0.02}_{-0.02}$&$\pm0.4^{2.3}_{-1.9}$&$\pm 0.03^{0.10}_{-0.12}$&$\pm 0.1^{0.2}_{-0.2}$\\
A2405  &$-23.38$&$ 93.9$&$-22.89$&$ 3.8$&   $-19.09$&H&8&[10778,11308]&$11020\pm70$&$140^{+80}_{-60}$\\
&$\pm 0.02^{0.25}_{-0.40}$&$\pm 4.7^{67.3}_{-33.5}$&$\pm 0.01^{0.06}_{-0.05}$&$\pm 0.1^{0.2}_{-0.2}$\\
A2571  &$-25.11$&$107.7$&$-23.89$&$ 5.4$&   $-21.49$&H&41&[31103,34517]&$32710\pm110$&$670^{+80}_{-70}$\\
&$\pm 0.01^{0.17}_{-0.22}$&$\pm 5.0^{43.5}_{-27.9}$&$\pm 0.02^{0.08}_{-0.09}$&$\pm 0.1^{0.4}_{-0.4}$\\
A2721  &$-25.02$&$ 47.9$&$-23.14$&$18.1$&     ...	 &N&67&[32452,36859]&$34300\pm100$&$840^{+80}_{-70}$\\
&$\pm 0.03^{0.04}_{-0.04}$&$ \pm 0.8^{1.8}_{-1.3}$&$\pm 0.14^{0.08}_{-0.08}$&$\pm 1.6^{1.2}_{-1.1}$\\
A2730  &$-26.00$&$245.9$&$-23.85$&$ 8.1$&   $-25.18$&N&13&[34281,37767]&$36000\pm300$&$1020^{+150}_{-140}$\\
&$\pm 0.03^{0.25}_{-0.32}$&$\pm13.5^{121.9}_{-78.4}$&$\pm 0.03^{0.19}_{-0.16}$&$\pm 0.2^{1.2}_{-1.3}$\\
A2811  &$-25.53$&$139.5$&$-23.59$&$ 5.5$&   $-24.72$&N&35&[30365,34227]&$32350\pm150$&$860^{+110}_{-100}$\\
&$\pm 0.03^{0.22}_{-0.23}$&$\pm 8.6^{59.7}_{-44.4}$&$\pm 0.03^{0.22}_{-0.15}$&$\pm 0.2^{0.8}_{-1.0}$\\
A2955  &$-25.22$&$ 93.5$&$-22.61$&$ 3.4$&   $-24.24$&H&22&[27539,29118]&$28260\pm70$&$320^{+90}_{-70}$\\
&$\pm 0.01^{0.08}_{-0.08}$&$\pm2.3^{11.3}_{-9.8}$&$\pm 0.03^{0.09}_{-0.08}$&$\pm 0.1^{0.3}_{-0.3}$   \\
A2969  &$-25.35$&$232.0$&$-23.90$&$19.7$ &   $-24.64$&N&21&[35684,40268]&$37740\pm220$&$980^{+190}_{-160}$\\
&$\pm 0.03^{0.27}_{-0.49}$&$\pm26.7^{203.6}_{-83.2}$&$\pm 0.10^{0.11}_{-0.06}$&$\pm 0.8^{1.8}_{-1.7}$\\
A2984  &$-25.79$&$202.7$&$-23.04$&$ 6.1$&   $-25.22$&H&29&[29882,32379]&$31230\pm90$&$490^{+110}_{-90}$\\
&$\pm 0.01^{0.01}_{-0.09}$&$ \pm5.4^{23.2}_{-2.4}$&$\pm 0.02^{0.01}_{-0.04}$&$\pm 0.1^{0.2}_{-0.0}$\\
A3112  &$-25.74$&$102.1$&$-22.95$&$ 8.3$&   $-23.79$&N&59&[19957,24560]&$22560\pm120$&$940^{+140}_{-120}$\\
&$\pm 0.01^{0.06}_{-0.06}$&$ \pm 1.7^{12.4}_{-9.4}$&$\pm 0.04^{0.16}_{-0.17}$&$\pm 0.2^{0.9}_{-0.7}$\\
A3166  &$-24.65$&$ 69.2$&$-21.92$&$ 1.7$&   $-21.73$&H&12&[34794,35511]&$35190\pm70$&$210^{+40}_{-40}$\\
&$\pm 0.02^{0.07}_{-0.08}$&$ \pm2.6^{8.6}_{-6.4}$&$\pm 0.04^{0.08}_{-0.08}$&$\pm 0.1^{0.2}_{-0.1}$\\
A3693  &$-25.46$&$314.5$&$-23.56$&$ 4.9$&   $-24.81$&H&31&[33500,38657]&$36840\pm190$&$1030^{+150}_{-130}$\\
&$\pm 0.05^{0.20}_{-0.21}$&$ \pm24.4^{61.5}_{-65.4}$&$\pm 0.02^{0.10}_{-0.04}$&$\pm 0.1^{0.3}_{-0.5}$\\
A3705  &$-24.27$&$ 30.8$&$-21.99$&$ 2.1$&   $-22.57$&N&53&[25110,29106]&$26820\pm140$&$1010^{+80}_{-80}$\\
&$\pm 0.01^{0.03}_{-0.01}$&$\pm 1.0^{10.9}_{-2.6}  $&$\pm 0.07^{0.02}_{-0.25}$&$\pm 0.1^{1.5}_{-0.3}$\\
A3727  &$-24.97$&$215.2$&$-23.07$&$ 6.1$&   $-24.16$&H&25&[33295,36030]&$34760\pm120$&$580^{+110}_{-100}$\\
&$\pm 0.05^{0.19}_{-0.21}$&$ \pm 17.1^{54.6}_{-43.7} $&$\pm 0.03^{0.05}_{-0.04}$&$\pm 0.1^{0.2}_{-0.3}$\\
A3809  &$-23.85$&$126.8$&$-24.14$&$20.8$&   $-22.79$&N&69&[17550,20192]&$18680\pm70$&$540^{+60}_{50}$\\
&$\pm 0.03^{0.13}_{-0.16}$&$ \pm 17.2^{31.4}_{-22.0}$&$\pm 0.03^{0.02}_{-0.02}$&$\pm 0.4^{0.1}_{-0.2}$\\
A3920  &$-24.80$&$ 45.6$&$-23.48$&$ 5.4$&	   ...	 &H&17&[37120,39232]&$38000\pm110$&$430^{+140}_{-100}$\\
&$\pm 0.01^{0.01}_{-0.04}$&$ \pm3.2^{4.4}_{-3.2}$&$\pm 0.10^{0.07}_{-0.05}$&$\pm 0.4^{0.1}_{-0.2}$\\
A4010  &$-25.56$&$186.3$&$-23.52$&$13.7$&   $-24.49$&N&31&[26957,30069]&$28620\pm120$&$630^{+150}_{-120}$\\
&$\pm 0.02^{0.19}_{-0.28}$&$ \pm10.7^{107.4}_{-58.1}$&$\pm 0.06^{0.37}_{-0.28}$&$\pm 0.6^{2.8}_{-3.0}$\\
APMC020&$-24.34$&$180.2$&$-24.40$&$17.6$&   $-23.41$&H&10&[32477,33228]&$32940\pm80$&$240^{+60}_{-50}$\\
&$\pm 0.04^{0.25}_{-0.62}$&$\pm34.4^{259.1}_{-71.3}$&$\pm 0.03^{0.07}_{-0.08}$&$\pm 0.4^{0.9}_{-0.7}$\\
AS0084 &$-25.33$&$ 72.0$&$-21.94$&$ 2.0$&   $-23.60$&N&24&[31436,34068]&$32390\pm120$&$520^{+160}_{-120}$\\
&$\pm 0.01^{0.03}_{-0.05}$&$\pm1.6^{5.4}_{-3.9}$&$\pm 0.04^{0.08}_{-0.10}$&$\pm 0.1^{0.2}_{-0.2}$\\
AS0296 &$-24.89$&$ 46.1$&$-23.99$&$10.0$&   $-20.36$&H&34&[20082,21872]&$20860\pm90$&$480^{+80}_{-70}$\\
&$\pm 0.01^{0.01}_{-0.01}$&$\pm 1.2^{7.5}_{-3.1}$&$\pm 0.04^{0.39}_{-0.12}$&$\pm 0.3^{0.9}_{-0.3}$\\
\enddata
\label{tab:data}
\end{deluxetable*}

The data were reduced in a standard manner, including bias subtraction using both
overscan regions and bias frames, and flat fielding using continuum lamp exposures
taken through the fibers. We used these same flat field exposures to define and trace the apertures,
allowing a spatial shift on the detector to best fit the position of the spectra in each individual exposure. We measured
redshifts using cross-correlation techniques, specifically the IRAF\footnote{IRAF is
distributed by the National Optical Observatories, which are operated by
AURA Inc., under contract to the NSF.} tasks within
the RVSAO package, XCSAO 
and EMSAO.
We visually inspected
all spectra both to reject marginal cases and to add systems with emission line redshifts that 
were missed by the cross-correlation. Most absorption line redshifts come from cross correlations
with $R > 4$. For spectra with both absorption and emission line redshifts, we give the average
of the two measurements. 
For the remaining 11 cluster fields from Paper I that were not targeted spectroscopically,
we obtained redshifts from the literature (using the NASA Extragalactic Database, NED) 
for galaxies projected within 1.5 Mpc of the BCG.

The kinematic results from our observations and literature search are given in Table \ref{tab:data}.
To determine cluster membership, we employ a pessimistic  3$\sigma$
clipping algorithm \citep{yv} using biweight estimators \citep{beers} of location and scale
for the cluster mean velocity and velocity dispersions, respectively \citep[cf.][]{zm}. The resulting 
number of cluster members, their velocity range, mean velocity, 
and line-of-sight velocity dispersion 
are presented in Table \ref{tab:data}, with the source entry referring to whether the
data come from our own Hydra (H) observations or from NED (N). Throughout the rest of this study,
we use the
line-of-sight velocity dispersion for the velocity dispersion of the CSphs,
assuming that the conversion factor is the same for all clusters and hence absorbed into the constant
term in $\log-\log$ relations (see \S3.2, 3.4, and 3.5).

Finally, we present 
the total luminosity of the resolved galaxies within the ICS $r_e$ in Table \ref{tab:data}.
We include this measurement because we will explore whether various relationships
are strengthened when using the entire luminous content of clusters.
We compute a background-subtracted luminosity within $r_e$, including all
galaxies fainter than the BCG down to $m=19.5$. We use the SExtractor \citep{sx}
stellarity index, which is robust down to this magnitude, to exclude
stars, and compute the background level using galaxies located at
distances of 23.3$\arcmin-46.6$\arcmin (2000-4000 pixels) from the BCG. 
We also apply a completeness correction to account for the contribution
of faint cluster galaxies at $m>19.5$. For this correction we adopt
$\alpha=-1.21$ and $M_{*,R}=-21.96$ \citep{cz} and
$R-I=0.82$ to convert from $R$ to $I$.

\section{Results and Discussion}

\label{sec:discussion}

\subsection{Effective Radius and Surface Brightness}

A key to the existence of the FP
is the inverse relationship between $r_e$ and $I_e$, the effective radius
and the mean surface brightness within that radius, respectively.  In Paper I, we demonstrate that
both the inner component of brightest cluster galaxies, which we associate with the BCG itself,
and the outer component, which we refer to as the intracluster light or 
intracluster stars (ICS),
exhibit a tight relationship between $r_e$ and $I_e$, although the slopes of
these relationships for the BCG and ICS components
differ slightly. We adopt the general practice of expressing $r_e$ in units
of kpc and $I_e$ in units of L$_{\odot}/$pc$^2$.

\begin{figure}
\plotone{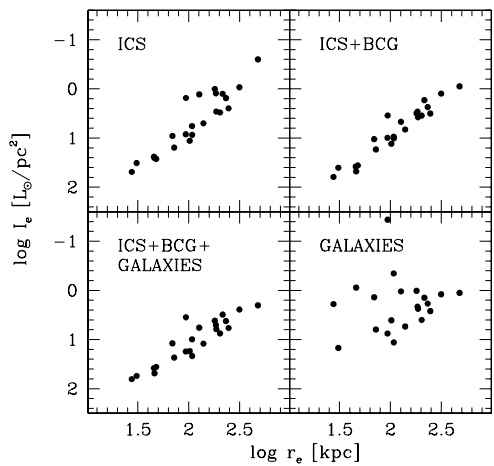}
\caption{The relationship between $r_e$ and $I_e$ for different definitions of $I_e$. 
In all panels we use $r_e$ measured from the ICS.
In the upper left panel we use
only the ICS luminosity within $r_e$ to calculate $I_e$. In the upper right panel, we add the
BCG luminosity to the calculation of $I_e$. In the lower left panel, we add both the BCG luminosity 
and that of the resolved cluster galaxies within $r_e$. In the lower right panel, we only use 
the luminosity of resolved cluster galaxies within $r_e$. }
\label{fig:reie}
\end{figure}

Because the luminosity within the ICS $r_e$ comes from at least three components
(BCG, ICS, and resolved galaxies), there is some ambiguity in deciding which luminosity to 
use in the calculation of $r_e$ and $I_e$ for the CSph.
In all four panels of Figure \ref{fig:reie}, 
we choose to plot $r_e$ for the ICS component, but alter our choice of the luminosity
used to calculate $I_e$. 
In the upper left panel, we include only the surface brightness of the ICS.
By construction for a $r^{1/4}$ profile, half of the total ICS luminosity
is enclosed within $r_e$.  In the upper right panel, we have added the entire BCG luminosity
to that of the ICS enclosed within $r_e$. The
addition of the two luminosities means that the adopted $r_e$  no longer
encloses half the total light. However, the BCG component is typically a small
fraction of the total light of the combined system (Paper I) and so does not 
drastically affect $r_e$, i.e., the half-light radius for the 
combined two components is typically 20 to 30\% smaller than the ICS $r_e$.
In the few cases where the BCG component dominates, the combined light
$r_e$ is several times smaller than the ICS $r_e$. 
Because 1) the ICS generally dominates,  2) $r_e$ and $I_e$ are correlated, and 
3) 
this correlation tends to slide the data along the $r_e-I_e$ relationship when $r_e$ changes, our use of the 
combined light $r_e$, instead of  the ICS $r_e$, does not significantly alter 
the slope or 
scatter of the various relationships we discuss.
Regardless of which $r_e$ is adopted,
including the BCG component in the calculation of $I_e$
decreases  the scatter.
Lastly, we 
include the luminosity from resolved cluster members within $r_e$ (lower left panel
of Figure \ref{fig:reie}). The empirical result, for which we will discuss a possible interpretation
below, is that the $r_e - I_e$ relation is tightest when we combine the luminosity
from the central BCG and the ICS to describe the CSph.   This result is supported
by a quantitative measure of the scatter and does not rest on a few deviant points.
When the BCG luminosity is included, the eight
clusters with a residual that is $> 0.1$ in the ICS $r_e-I_e$ correlation all drop to below $0.1$
and no clusters increase their residual to $> 0.1$.

\begin{figure}
\plotone{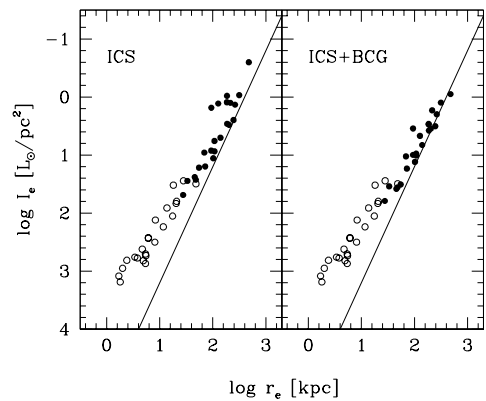}
\caption{The relationship between $r_e$ and $I_e$ for both ICSs and BCGs. 
In both panels the open circles
represent the data for the BCGs. In the left panel, the filled circles represent the data
for the ICS. In the right panel, the filled circles represent the calculated $I_e$ using the 
sum of the 
ICS and BCG luminosities within the ICS $r_e$. The solid line illustrates the direction along
which errors in the measurement of $r_e$ will scatter the data.}
\label{fig:rescatter}
\end{figure}

A difficulty in interpreting the $r_e-I_e$ relationship is that $I_e$ depends directly on
the determination of $r_e$. This degeneracy does not render the relationship
meaningless because there is additional information in $I_e$ regarding the luminosity profile. 
However, the 
degeneracy does imply that errors in $r_e$ will translate into correlated
errors in $I_e$. The observed relationship is
close in slope to that arising from correlated errors and is therefore 
somewhat suspect (Figure \ref{fig:rescatter}). At
this point one must either have confidence in the quoted errors
or provide independent measurements demonstrating that 
$r_e$ and $I_e$ are not spurious.
Although our uncertainties are indeed sufficiently small that we claim 
not to have confused a large $r_e$ ICS component with a small one (see the original
Figure 9 in Paper I or Table \ref{tab:data}), these are difficult measurements
that are susceptible to systematic errors, particularly in the sky determination (see
Paper I). Given both the random and systematic errors, we 
cannot easily rule out the possibility that some luminosity is artificially taken from one component
and placed into the other by our fitting.

We first examine whether another measurement can indirectly validate our measurement
of $r_e$.  For example, our measurement of
$r_e$ for the ICS correlates with the velocity dispersion of the cluster, $\sigma$,
sufficiently well that
we can rule out at the 97\% 
confidence level that such a correlation arose randomly.
This correlation suggests that there is real information
in the $r_e$ measurements (a lack of a correlation would not conversely suggest
that the $r_e$ values were dominated by errors).
In contrast to this ICS result, the $r_e$ values for the BCG
correlate only weakly with $\sigma$ and there is a 13\% chance that such
a correlation could arise at random. 

\begin{figure}
\plotone{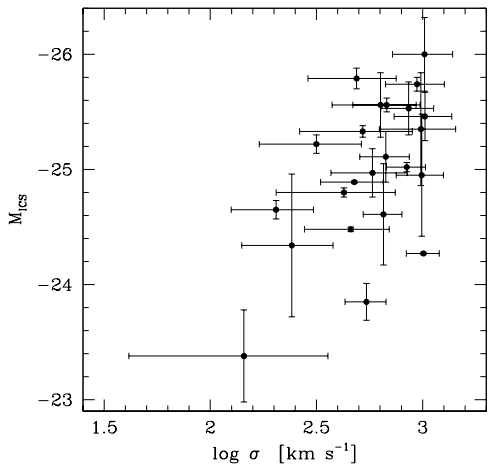}
\caption{Cluster velocity dispersion versus ICS magnitude. Errorbars in $M_{ICS}$ represent
the potential systematic error due to a $1\sigma$ systematic sky error. The formal
fitting errors in $M_{ICS}$ are significantly smaller than the plotted systematic errors.}
\label{fig:vsig}
\end{figure}

Next we examine whether the measurement of the ICS luminosity correlates
with cluster velocity dispersion. We find again that there is a strong correlation (98\% confidence
level). Given the large uncertainties in $\sigma$ (Figure \ref{fig:vsig}), the underlying
correlation must be significantly tighter, which once again suggests that errors in the
sky determination are not sufficiently large to disrupt our measurements of the ICS.
As with $r_e$, the correlation for the BCG component
is much weaker (a 20\% chance of random).

\begin{figure}
\plotone{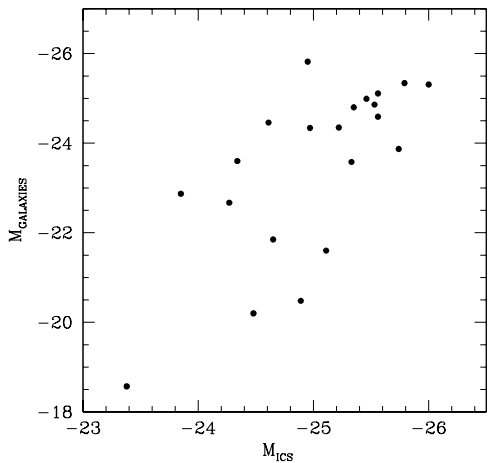}
\caption{Magnitude of summed cluster galaxies inside $r_e$ versus the magnitude of the ICS.}
\label{fig:lplot}
\end{figure}

Finally, we compare the ICS magnitude
to that of  the summed cluster galaxies inside $r_e$ (Figure \ref{fig:lplot}).
We find a strong correlation
(99.97\% confidence level) between these two quantities, 
and again the correlation is much weaker between the BCG and
the resolved galaxy luminosity (27\% chance of random).
We conclude that our measurements of
$r_e$ and $L_e$, which in turn define $I_e$,  for the ICS
are reliable and that the range of these values does have physical meaning.
Nevertheless, when deriving quantitative measurements
we take care to account for the correlated
errors through Monte-Carlo simulations.

We conclude from these tests that our structural measurements, particularly those
for the ICS, accurately reflect the properties of the cluster. We are left with two families 
of interpretations for the measured improvement in the $r_e-I_e$ relation when the
combination of the BCG and ICS luminosity is used. The 
improvement may simply arise because our partition of luminosity between the BCG 
and ICS is poor, and therefore the correlation using the sum of components 
has less scatter than that using one poorly determined component. Alternatively,
the improvement may indicate some deeper connection between the BCG and ICS
and that the two together properly reflect the CSph.  We favor the latter because
(1) adding the BCG luminosity reduces
the scatter both for systems with too little and too much luminosity within
$r_e$, (2) there is 
no relationship between the scatter in the $r_e -I_e$ relation and the estimated
systematic error, and (3) the decomposition of components in Paper I was based
not solely on a surface brightness profile, but also on ellipticity and position angle
variations.  We also favor the latter because there are external arguments
for a connection between the properties of BCGs and the global cluster environment. 
These connections include the finding that 
BCGs are not drawn from the luminosity function of other cluster ellipticals \citep{tr, d78}, and that 
there is a correspondence between the BCG optical  luminosity and the 
cluster X-ray luminosity \citep{hudson,burke, nelson}. In the final analysis, none of these
arguments is definitive, and one could prefer to simply fall back to the empirical improvement
in the scatter to adopt the joint BCG and ICS luminosity. While we choose
to use that combination as the best description of the CSph, we also present  results using 
only the ICS luminosity, and show that 
this choice does not materially affect any of our findings and conclusions.
Unless otherwise noted, from now on we use the sum of the BCG
and ICS components for the CSph.

\subsection{Defining the CSph Fundamental Plane}

Emboldened by the strong $r_e-I_e$ relationship for the CSphs, and following the 
example of the FP for elliptical galaxies, we examine whether the cluster velocity
dispersion helps further refine the portion of parameter space
occupied by our cluster spheroids. While
the most appropriate analog would be the velocity dispersion of the ICS component itself,
the low surface brightness of the ICS precludes measurement of $\sigma$ in all
but a few clusters (see \cite{kelson}).   We showed above that various properties
of the ICS are correlated with the cluster velocity dispersion as measured from
the galaxies, and so the use of the cluster velocity dispersion as a proxy for
the dispersion of the ICS is not without reason. 
The data are so far inconclusive on the question of whether this is a good assumption.
In one cluster the ICS dispersion is 
consistent with that measured from galaxies at large radius \citep{kelson}, and in
another it is not \citep{gerhard}. 
A further departure of our work from a direct analogy with
the study of ellipticals is that the velocity dispersions of ellipticals are measured within $r_e$, or
corrected to represent the dispersion within $r_e$.
Because there are typically few cluster galaxies within $r_e$, we
adopt the global cluster velocity dispersion as our velocity measurement with no
correction.
All of these departures should weaken any underlying CSph FP relation, and
as such our results will be conservative.

\begin{figure}
\plotone{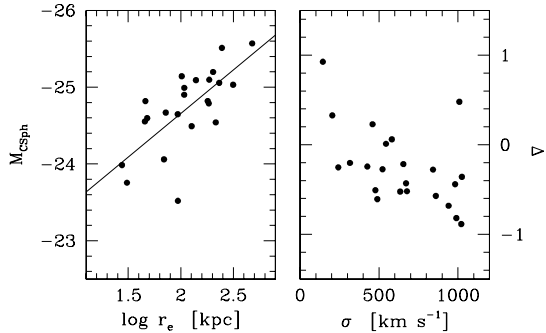}
\caption{Connecting $r_e$, $L$, and $\sigma$ for the CSphs.
Left panel shows the correlation between ICS $r_e$ and
CSph (ICS+BCG) magnitude. The solid line represents the best fit line to those data.
The right panel shows the correlation between the cluster velocity dispersion $\sigma$ 
and the residuals to the best
fit line from the left panel.}
\label{fig:lplot1}
\end{figure}

Before proceeding to the FP analysis, we ask 
if $\sigma$ is correlated with the residuals of any of the correlations
examined so far.  If so, such a correlation would further suggest that the cluster
velocity dispersion is related to the structural properties of the CSph and provide
additional motivation to proceed with an examination of the FP. In Figure \ref{fig:lplot1},
we show the $r_e-L_e$ relationship,
and the residuals about that relationship, $\Delta$, versus $\sigma$. 
We examine residuals about the $r_e-L_e$ relationship rather
than about  the $r_e-I_e$ relationship because $r_e$ and $L_e$ ($\equiv L/2$) 
are independently measured quantities.
The $\Delta-\sigma$ relationship is highly significant 
(99.2\% confidence level)
and demonstrates that
the velocity dispersion does provide additional information on the structure of
these systems. 

The strong relation between the ICS effective radius 
and the combined mean surface brightness of the CSph (upper right
panel of Figure \ref{fig:reie},
with a scatter of 0.15 in the plotted units) already suggests that the principal extension of the CSph fundamental
plane, if there is one,  is primarily along these two axes. 
As is generally done for ellipticals, and to facilitate comparison with them, we include $\sigma$ as the third 
axis and fit 
$$A \log{\sigma} + B \log{I_e} - \log{r_e} + C = 0. \eqno{(1)}$$

The fitting of a plane to these data is complicated for several
reasons.  First, in a
typical fitting problem,  the independent variables are much less
uncertain than the dependent variables, and hence errors in the independent variables can
be safely ignored. In this case, however,
not only are the uncertainties comparable for 
all three variables, but the uncertainties in two of these variables are highly correlated.
Second, there are multiple ways in which one can define residuals about a plane.
In the direct method (for a more detailed discussion of the nomenclature see 
\cite{bernardi}), one minimizes the 
residuals along an axis (typically, residuals in $r_e$ are quoted), while in the
orthogonal method one minimizes the 3-D distance between the data 
and the plane. These two methods appear to give roughly the same answer for
$B$ in Eq. 1 when fitting Es, but differ by about 25\% in their determination of $A$ 
in Eq. 1 \citep{bernardi}. Although
the choice of method is important 
for quantitative measurements, the prime consideration when comparing results
from various studies is whether the same method is used throughout.
We choose to focus on results from the direct method for continuity throughout the
paper and for the technical reasons that develop below.
Therefore, we will
refit some previously published data that were fit originally with the orthogonal method.
For our exploration of the CSph FP alone,
we present both direct and orthogonal fits in Table \ref{tab:fp}.

\begin{figure}
\plotone{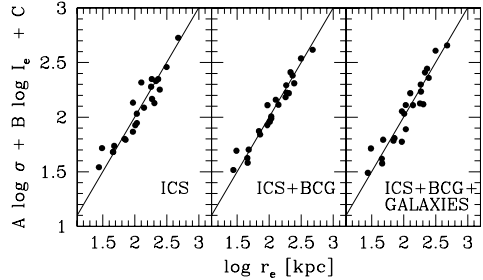}
\caption{Fundamental Plane for CSphs. We show projections of the FP for the
ICS only (left panel), ICS+BCG (middle panel), and ICS+BCG+Galaxies (right panel) components.
The line is the 1:1 line, which is the plane in these appropriately scaled axes.
The numerical values for $A, B$, and $C$ are given in Table \ref{tab:fp}.
\label{fig:fp}}
\end{figure}

Although the fitting problem has been solved analytically for Gaussian errors 
\citep[see][]{bernardi}, we
opt for a numerical fitting method so that we can experiment with 
asymmetric error distribution, and eventually move beyond planar
fits in a straightforward way (\S3.4). We fit a plane to our
data, using $r_e$ as measured from the ICS surface brightness profile, $I_e$ as calculated
from the adopted luminosity profile (ICS, ICS+BCG, or ICS+BCG+galaxies, see below), and  $\sigma$ as measured
from member galaxies. To account for measurement uncertainties, and the correlated
errors of $r_e$ and $I_e$, we randomly draw errors for $r_e$, $L$, and $\sigma$ from 
the uncertainties given in Table \ref{tab:data} and then recalculate $L_e$ and $I_e$. 
We use the estimated systematic uncertainties resulting from a 1$\sigma$ global error in the
background sky determination because these are larger than the statistical
errors for the measured parameters.
Where errors are asymmetric, we adopt Gaussians of different 
dispersions in the two directions. 
We repeat the fitting 1000 times and then determine uncertainties in the fitted
parameters by identifying the range that encloses two-thirds
of the results from the 1000 trials. These fits, particularly in cases where the
errors are asymmetric, will not result in the same fitted parameters as a direct fit to
the original data because the mean value of the
distribution is not equal to the observed value. In general, the resulting difference
between the fitted parameters is undetectable.

\begin{deluxetable*}{llrrrr}
\tablecaption{CSph Fundamental Plane Coefficients}
\label{tab1}
\tablewidth{0pt}
\tablehead{
\colhead{Method} & 
\colhead{Components} & 
\colhead{$A$}   & 
\colhead{$B$}   &
\colhead{$C$}   &
\colhead{$r_e$ RMS}   
\\
}
\startdata

Direct \\
\\
&ICS&0.26$^{+0.06}_{-0.08}$&$-0.48^{+0.02}_{-0.02}$&1.66$^{+0.22}_{-0.18}$&0.106\\
&ICS+BCG &0.21$^{+0.02}_{-0.02}$&$-0.56^{+0.01}_{-0.01}$&1.96$^{+0.05}_{-0.06}$&0.074\\
&ICS+BCG+Galaxies & 0.32$^{+0.06}_{-0.07}$&$-0.65^{+0.03}_{-0.03}$&1.86$^{+0.21}_{-0.17}$&0.101\\
\\
Orthogonal\\
\\
&ICS&0.47$^{+0.07}_{-0.17}$&$-0.51^{+0.03}_{-0.03}$&1.10$^{+0.51}_{-0.18}$&0.112\\
&ICS+BCG &0.23$^{+0.02}_{-0.04}$&$-0.57^{+0.01}_{-0.01}$&1.91$^{+0.06}_{-0.07}$&0.073\\
&ICS+BCG+Galaxies & 0.49$^{+0.05}_{-0.14}$&$-0.71^{+0.03}_{-0.03}$&1.44$^{+0.40}_{-0.13}$&0.107\\
\\
\enddata
\label{tab:fp}
\end{deluxetable*}

In Figure \ref{fig:fp} we plot edge-on projections of the fitted plane and data 
for the various plausible choices of the CSph luminosity. 
Again, as in \S3.1, we explore adopting either the luminosity from
just the ICS, from the combination of the BCG and ICS, or from the combination
of the BCG, ICS, and the resolved cluster members within $r_e$.
The scatter about the plane is 0.106 when we use only the ICS
luminosity to determine $I_e$, 0.074 when we use 
the BCG+ICS luminosity, and 0.101 when we also include the resolved galaxies
within $r_e$.  Once again the combined BCG+ICS luminosity provides the
tightest relationship\footnote{
We confirm that the reduction in the scatter when the
BCG light is included is statistically significant
by fitting 1000 randomized versions of the
data. We draw the relevant parameters from Gaussian distributions
according to their estimated uncertainties. In this manner we are
able to answer the questions: 1) if the tight correlation seen in
the ICS+BCG data is correct, what is the likelihood of getting as
``poor'' a correlation as that seen for the ICS data alone by random chance,
and 2) if the noisier correlation seen in the ICS data alone is
correct, what is the likelihood of getting as ``tight'' a correlation
as that seen for the ICS+BCG data by random chance.
From these Monte-Carlo
simulations, we find that none of the 1000
trials of the ICS+BCG data produces as large a $\chi^2$ value as that
of the fit to the ICS data alone and that none of the 1000 trials of the
ICS data alone produces as small a $\chi^2$ as that measured for the ICS+BCG data.}. 
Therefore,  we conclude that the CSph is best
described by the BCG+ICS and adopt this description for the CSph unless otherwise noted.
None of the basic results we present depends 
on which definition is used (see Table \ref{tab:fp}).
The change in the scatter between the 0.15 measured
for the $r_e-I_e$ relationship and the 0.074 measured for the CSph FP demonstrates
the improvement introduced by utilizing the velocity dispersion. The low scatter relative
to the FPs of other spheroids
suggests that using the velocity dispersion of the ICS rather than that of the galaxies
is unlikely to significantly improve the relation. This result does not necessarily
imply 
that the velocity dispersion of the ICS is the same as that of the cluster
galaxies, but rather that
they scale proportionally so that the scaling between the two is absorbed into 
the constant term in the FP equation.
The fitted parameters and rms scatter between the fitted and observed $\log r_e$ values are
listed in Table \ref{tab:fp}.

\subsection{Comparison to Other Spheroids}

The existence of a FP for CSphs leads naturally to the question of whether this
FP is similar to that found for other, smaller spheroids. Due to a wide range of
studies, data exist for systems as diverse as dSphs,  dEs, Es, and BCGs. Of these,
the most extensive studies of the FP have been done with normal  Es, so we begin
with a comparison to those galaxies, then compare to other spheroids, and finally attempt
to place all spheroids on a single relationship.

The degree of scatter  in the CSph FP 
is slightly lower than that typically obtained for elliptical galaxies
(see \cite{jorg,bernardi}). To enable a more direct comparison, we apply our direct fitting 
technique to the elliptical data compiled by \cite{jorg}.
The direct fitted parameters ($A = 1.00$, $B = -0.75$, and a
scatter of 0.094)
differ slightly from those derived 
by \cite{jorg} using the orthogonal fitting method ($A = 1.24$, $B= -0.82$ and 
a scatter of 0.084), but we recover those published values when we fit
the data using the orthogonal approach. 

Both the direct and orthogonal fits  to the elliptical galaxy data differ significantly
from the corresponding fits to the CSphs, and that difference is
primarily in the value of $A$. 
Therefore, while the CSphs
fall onto a FP as tightly as do elliptical galaxies, it is not the same plane.
This comparison, however, is complicated by the use of data obtained with 
different filters (for example, $r$ from \cite{jorg} and $I_C$ from Paper I). We correct the \cite{jorg}
data to $I_C$, hereafter $I$, using a model color calculation for Es by \cite{fukugita}, but 
there remains some discrepancy between the two samples
that depends on the degree to which (1) $r_e$ is color dependent (see \cite{pahre} for
evidence of color dependence)
and (2) the colors of all ellipticals differ from the models.
Nevertheless, these uncertainties will 
only increase the scatter in any attempt to place the full range of spheroidals
on one relationship.

We fit  a plane to the BCG data of \cite{oh}, again correcting using the 
colors of elliptical galaxies and the calculations by \cite{fukugita}. In this case, we 
find a planar fit that has an $A$ value of 0.52,  which is between those of the ellipticals (1.00 or 1.21)
and CSphs (0.21).  The BCG scatter about the best-fit plane  (0.085)
is comparable to that found for both Es and CSphs. These 
results suggest that each of 
these distinct types of spheroids (Es, BCGs, and CSphs) lies on a comparably thin,
yet distinct,  plane in this space.

\begin{figure}
\plotone{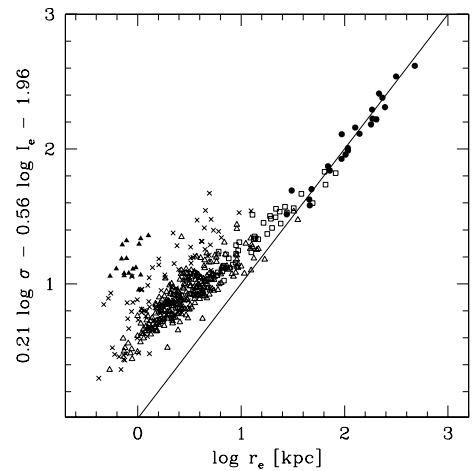}
\caption{The projection of other spheroid populations onto the CSph FP. The
extrapolation of the CSph FP to lower $\sigma$ is illustrated with the solid line.
Solid circles represent our CSph measurements, open
squares the BCGs of \cite{oh}, open triangles the Es of \cite{jorg}, crosses the Es 
of \cite{coma}, and solid triangles the dEs of  \cite{geha}. There is no single FP for
all spheroids.}
\label{fig:fpall1}
\end{figure}

One way of comparing the various spheroidal populations is to project them onto 
the CSph FP (Figure \ref{fig:fpall1}). We infer from the
continuity of the various samples in this plot that a single
relationship may unite all systems from Es to
CSphs. This
underlying unifying relationship is more complex than a simple plane.
The trend of decreasing $A$ from ellipticals to BCGs to CSphs
suggests that rather than three distinct planes, these spheroids may populate
a slightly twisted 2-D manifold. To extend the range of spheroids, we also plot
several other samples that contain lower $\sigma$ systems \citep{coma, geha},
again correcting the photometry using either
model colors and the tabulations by \cite{fukugita} or measured colors (for the \cite{geha} sample,
we use the mean measured colors for dEs from \cite{stiavelli}). Although the
existence of a single, unifying manifold is less obvious for these systems because they 
deviate from the E-BCG-CSph trend, we show below that they can be included.

\begin{figure}
\plotone{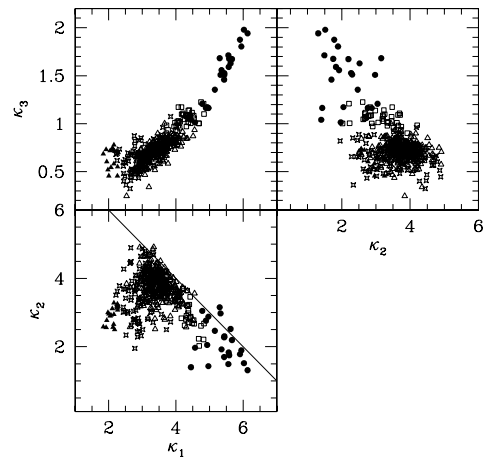}
\caption{Ellipticals, BCGs, and CSphs plotted in $\kappa$-space. 
Solid circles represent our CSph measurements, open
squares the BCGs of \cite{oh}, open triangles the Es of \cite{jorg}, crosses the Es 
of \cite{coma}, and solid triangles the dEs of  \cite{geha}. The line in the lower panel
marks the boundary identified by \cite{burstein}.
\label{fig:kappa}}
\end{figure}

We conclude this section by placing these systems in $\kappa$-space \citep{bender}. 
The CSph data appear as an extension of the
elliptical and BCG data in all three projections, again suggesting a physical connection 
among spheroids of different scales (Figure
\ref{fig:kappa}).
In the $\kappa_1 - \kappa_3$ projection,
which corresponds to the mass$ - \log (M_e/L_e)$ projection, the CSphs map onto the
distribution of Es and extend that relationship up to 
the very large $M/L$ ratios of clusters. The tightness of the trend in this panel
suggests that much of the difference among spheroids reflects 
differences in the relative importance of dark matter within $r_e$. Previous investigators
of the tilt of the FP \citep{faber, ps96, trujillo} have considered the effect of systematic
changes in $M_e/L_e$ with other galaxy properties and reached varying conclusions.
The advantage provided by comparing our data to that of smaller spheroids is that the
range of $M_e/L_e$ is so large that some causes of $M/L$ differences, such as stellar age and
metallicity variations, are not physically plausible explanations for the entire relationship. 
Variations in $M/L$ with galaxy mass were previously identified \citep[cf.][]{burstein}, but our
analysis is different in its use of the properties of the CSph, rather than the member galaxies
\citep{schaeffer}, and in the 
continuity and overlap among those 
samples, which enables cross-checking for systematic differences
among samples. These advantages will enable us to identify a simple relationship between $M_e/L_e$
and $\sigma$.


\subsection{Beyond the Fundamental Plane}

The standard, back-of-the-envelope derivation of the FP begins with the virial
theorem. Subsequent non-trivial simplifications, such as the use of $\sigma$ 
in the kinetic energy term, the constancy of the numerical conversion factor between the
measured line-of-sight velocity dispersion and the 3-D velocity dispersion
$\sigma$ for all spheroids, the use of 
$r_e$ in the potential energy term, the self-similarity of the potentials for all spheroids, as well as the assumption of sphericity, produce
$$\sigma^2 \propto M_e/r_e,\eqno{(2)}$$
where the proportionality constant depends on details of the structure that are
assumed to be consistent among spheroids. The assumption that the structure of spheroids
is scaleable and has no preferred size is referred to as homology.
By converting mass to $(M/L)L$, Eq. 2 is rewritten as
$$\sigma^2 \propto (M_e/L_e)(I_er_e^2)/r_e. \eqno{(3)}$$
Taking the logarithm and rearranging terms, we arrive at
$$\log r_e = 2 \log \sigma - \log I_e - \log (M_e/L_e) + C, \eqno{(4)}$$
which is the source of the expectation that the FP coefficients $A$ and $B$
should be 2 and 1, respectively (see Eq. 1). The ``tilt" in the FP, which corresponds to 
$A$ and $B$ not having the values this simple calculation suggests, is often attributed
to the neglected $M_e/L_e$ term, or to the breakdown of the various simplifying assumptions, 
principally homology.

\begin{deluxetable}{lrrrr}
\tablecaption{Direct Fundamental Plane Fits For Different Spheroids}
\label{tab1}
\tablewidth{0pt}
\tablehead{
\colhead{Components}& 
\colhead{$A$}& 
\colhead{$B$}&
\colhead{$r_e$ RMS}   
}
\startdata
Es\citep{jorg}&$1.00^{+0.01}_{-0.01}$&$-0.75^{+0.005}_{-0.003}$&0.094\\

Es\citep{bernardi}&$1.21^{+0.04}_{-0.04}$&$-0.77^{+0.01}_{-0.01}$&0.085\\
BCG\citep{oh}&$0.52^{+0.09}_{-0.10}$&$-0.87^{+0.003}_{-0.002}$&0.089\\
CSph&$0.21^{+0.02}_{-0.02}$&$-0.56^{+0.01}_{-0.01}$&0.074\\
\enddata
\label{tab:fpcomp}
\end{deluxetable}

One difficulty in addressing whether Eq. 4 is an accurate description of 
spheroids is that there is no {\sl a priori} expectation for how $M_e/L_e$ varies with
any of the other parameters. That dependence has generally been 
solved for after the fact by requiring it to reproduce the observed values of $A$ and $B$ 
\citep{faber,ps96}. 
This approach is underconstrained because a solution is guaranteed if a tilted
plane originally fit the data. In other words, the reverse engineering of
$M_e/L_e$
is telling us either something about how it varies with galaxy properties or that
the simple formulation of the FP in Eq. 4 has failed.
External checks, such as a comparison between the inferred $M_e/L_e$ and the colors of
the stellar population, suggest that there is at least some dependence between
$L$ and $M_e/L_e$. However,  a wider range of data, including $K$ band photometry, which
is less sensitive to variations in stellar populations, demonstrates that only
a small fraction of the overall tilt observed in the elliptical FP can come from
stellar population differences \citep{pahre, trujillo}. Because the 
change in $M_e/L_e$ across the range of normal elliptical galaxies is modest (typically a factor of a few at most),
there are several different physical causes that could give rise to the FP tilt.
Observing a sample that spans a much larger range of luminosities or velocity dispersions
provides a critical test of the inferences drawn
from normal ellipticals alone \citep[see][]{napolitano}.

\begin{figure}
\plotone{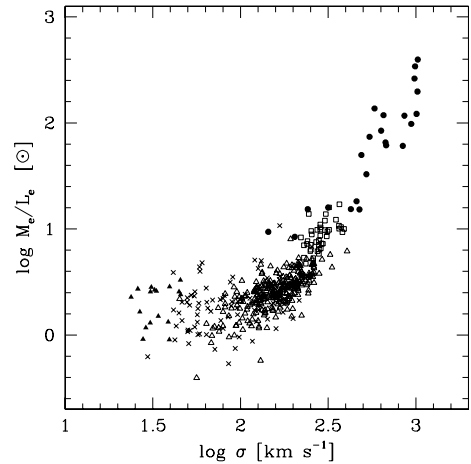}
\caption{$M_e/L_e$ versus $\sigma$. $M_e/L_e$ is calculated using $\kappa_3$ and
is in solar units corresponding to the $I-$band
(data other than ours is corrected to this band; see text). Because these
values are calculated assuming homology, they are not internally consistent
with our finding that $M_e/L_e$ does not scale as a power law of $\sigma$ and
may not represent in detail the true relationship between  $M_e/L_e$ and $\sigma$.
Nevertheless, the qualitative result of a non-linear relation between $M_e/L_e$ 
and $\sigma$ is robust.
Solid circles represent our CSph measurements, open
squares the BCGs of \cite{oh}, open triangles the Es of \cite{jorg}, crosses the Es 
of \cite{coma}, and solid triangles the dEs of  \cite{geha}. }
\label{fig:mtol1}
\end{figure}

On the basis of the trend in $A$ values seen in Table \ref{tab:fpcomp} and the tightness
of the sequence between $\kappa1$ and $\kappa3$,  we explore the
connection between $\sigma$ and $M_e/L_e$. We estimate $M_e/L_e$ using
a dimensional argument \citep[$M/L \propto \sigma^2/(I_er_e)$;][]{burstein} and plot the relationship between $M_e/L_e$ and $\sigma$
in Figure \ref{fig:mtol1}. It is evident that the relationship between these two
quantities is not a simple power law, as would be the case if the homology assumption
were correct. The range
in ${M_e/L_e}$ over which the relationship is now evident rules out earlier interpretations
of apparently linear trends, seen only over a modest range of $M_e/L_e$, 
as arising from stellar population differences. While
it is  possible that stellar population differences account for some of the 
$M_e/L_e$ variation, particularly for systems with small $M_e/L_e$,
they cannot account for the factor of several hundred
change in $M_e/L_e$ seen here. The same can be said about suggested causes of
homology breaking such as orbital anisotropy differences or systematic differences in
the surface brightness profiles.

The variations seen here, if attributable primarily to one cause, appear to reflect
the efficiency of concentrating and converting baryons to stars within $r_e$. 
The inability of a power law to describe the relationship between $M_e/L_e$  and $\sigma$
is clear from Figure \ref{fig:mtol1}, but it can also be easily demonstrated
that no simple power law dependence of $M_e/L_e$ on $\sigma$ can reproduce
the variation in the FP $A$ coefficient seen in \S3.3. 
Consider $M_e/L_e \propto \sigma^\alpha$. Such a relationship changes Eq. 4 to
$$\log r_e = (2 - \alpha)\log\sigma - \log I_e + C_1. \eqno{(5)}$$
While such a power law relationship could explain why FP fits do not find $A = 2$, it would
not explain why $A$ varies with $\sigma$. Similarly, postulating power law
dependences of $M_e/L_e$ on $I_e$ and $r_e$ would change the numerical value
of the coefficients in the FP, but not the functional form. Therefore, the 
description of $M_e/L_e$ as $\sigma^2/I_er_e$ cannot account for the varying tilts of
the FPs.

Instead, from an inspection of Figure \ref{fig:mtol1}, we  suggest a new
relationship:
$$\log M_e/L_e = (\alpha\log\sigma - \beta)^2 + \gamma, \eqno(6)$$
where $\alpha, \beta,$ and $\gamma$ are free parameters.
Such a relation 
modifies the original FP equation (Eq. 4) to
$$\log r_e = (2 - \alpha^2\log\sigma + 2\alpha\beta)\log\sigma - \log I_e + C_2 \eqno{(7)}$$
and results in an effective $A$ coefficient that
is a function of $\sigma$. One can imagine invoking other functional forms for $M_e/L_e(\sigma)$, 
but an offset parabola, a 2nd order polynomial, is the simplest 
mathematical form that can fit the nonlinear relationship in Figure \ref{fig:mtol1}. 
This form, which describes a twisting plane or a 2-D manifold, 
is not a unique choice, but we will show that the quality of the fit leaves little room
for improvement with alternative fitting functions.

\subsection{Relationship between $M_e/L_e$ and Velocity Dispersion}

We fit a parabola to the $M_e/L_e$ data for dwarf ellipticals, ellipticals, BCGs, and 
CSphs in Figure \ref{fig:mtol1}.  It is important to clarify two issues before proceeding. First, the $M_e/L_e$ values
that we fit are not derived from detailed dynamical modeling, but are rather those
calculated from a dimensional argument. 
As such they cannot be the correct values in
detail because, as we determined above, $M_e/L_e$ cannot be described as 
a simple ratio of $\sigma, I_e,$ and $r_e$ and produce the correct scaling of the $A$ coefficient. Nevertheless, these  $M_e/L_e$ values should roughly track the true
values because deviations from a single FP fit to all of these spheroids are not gross 
(rms scatter $\sim$ 0.2, see below).
Second, these values of  $M_e/L_e$, as well as those necessary to fit Eq. 7,
reflect the ratio within $r_e$, not the global (total) $M/L$.

\begin{deluxetable}{rrrc}
\tablecaption{Measurements Used in Fitting the Fundamental Manifold\tablenotemark{a}}
%\label{tab4}
%\tablewidth{0pt}
\tablehead{
\colhead{$\log r_e$}&
\colhead{$\log \sigma$}&
\colhead{$\log I_e$}&
\colhead{References\tablenotemark{b}}\\
\colhead{[kpc]} &
\colhead{[km s$^{-1}$]}&
\colhead{[$L_\odot  {\rm pc}^{-2}$]}& \\
}
\startdata
2.03&2.83&1.01&1,2\\
2.68&3.00&$-$0.05&1,2\\
2.27&2.82&0.46&1,2\\
1.44&2.66&1.79&1,2\\
1.97&2.16&0.54&1,2\\
2.03&2.83&0.98&1,2\\
1.68&2.93&1.56&1,2\\
\enddata
%
%\vskip 18pt
\noindent
\tablenotetext{a} {NOTE: The complete version of this table is in the electronic edition of the Journal.
The printed edition contains only a sample.}
\tablenotetext{b}{References: (1) This paper, (2) \cite{GZZ}, (3) \cite{jorg}, (4) \cite{oh}, (5) \cite{geha},
and (6) \cite{coma}.}
\label{tab:mdata}
\end{deluxetable}

\begin{figure*}
\plotone{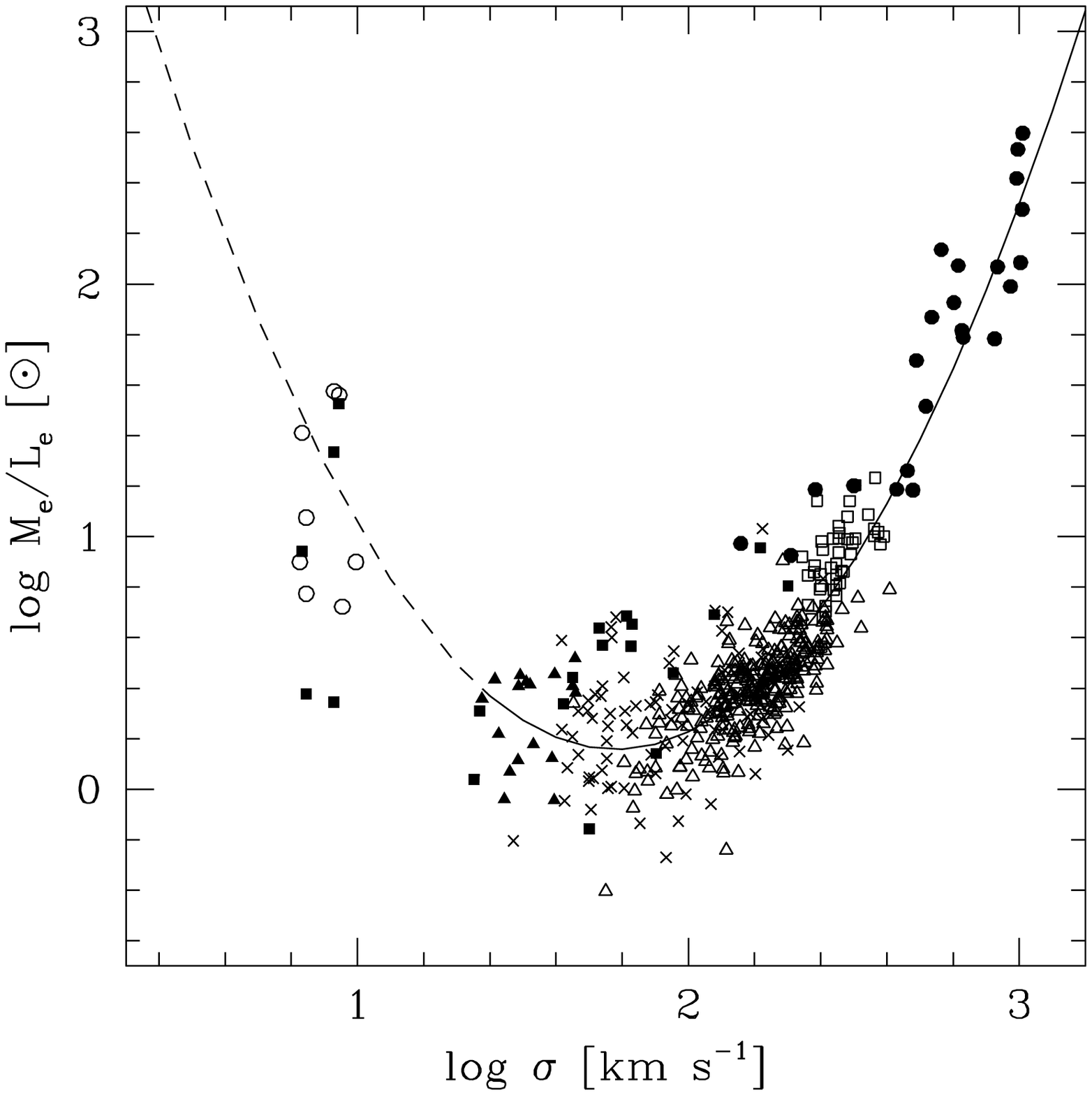}
\caption{An extended view of $M_e/L_e$ versus $\sigma$. To the data presented in Figure \ref{fig:mtol1},
we have added  dEs and dSphs from \cite{bender} (filled squares) and
dSphs (open circles) from \cite{mateo}. Five of the dSphs, at $\log \sigma \sim 1$, appear in both the \cite{bender}
and \cite{mateo} compilations, although the $M_e/L_e$ values are calculated in independent
ways. For the \cite{mateo} data, we adopt the lower $M_e/L_e$ value if two are given. 
We also add a curve illustrating the best-fit parabola. We use a dashed line for $\log\sigma < 1.4$
because the fit is unconstrained for these values. Neither the
\cite{bender} nor \cite{mateo} data are used in the fitting due to the large, possibly systematic, uncertainties. The simple parabola fits the data for systems ranging from dSphs to 
rich galaxy clusters.
}
\label{fig:mtol2}
\end{figure*}

In Figure \ref{fig:mtol2} we plot the same elliptical, BCG, and CSph $M_e/L_e$ values versus $\sigma$ 
as in Figure \ref{fig:mtol1}, but with several key additions. First, we have added
the fit to the data in Figure \ref{fig:mtol1} (solid line for the region constrained by data, dashed for the extrapolation). 
Second, we
have added the data for dwarf ellipticals and dwarf spheroidals from \cite{bender}
and for dwarf spheroidals from \cite{mateo}. We have corrected both data sets to the $I$-band
using the models of \cite{fukugita} and colors of dEs \citep{stiavelli} where necessary,
and corrected the  \cite{bender} data
using the currently accepted distances to the Local Group dSphs and our adopted 
value of $H_0$ for galaxies outside the Local Group. These resolved systems were not included
in the fitting, or in Figure \ref{fig:mtol1}, because their properties ($I_e$, $r_e$ and $M_e/L_e$) 
are measured so differently than for the other systems that systematic errors
in the comparison are likely. Even so, their properties
are in concordance with the extrapolated parabolic fit to the properties of the unresolved systems.
While there is large scatter in the $M_e/L_e$ measurements at the lowest $\sigma$'s, the
data support our suggestion
of a parabolic description for the relationship between $\log M_e/L_e$ and $\log\sigma$.
The detailed fit cannot be used to quantitatively formulate the ``fundamental manifold"
because the underlying assumptions are not internally self-consistent.
We exclude globular clusters from this discussion because they are not spheroids embedded
in a dark matter potential well. 

\begin{figure*}
\plotone{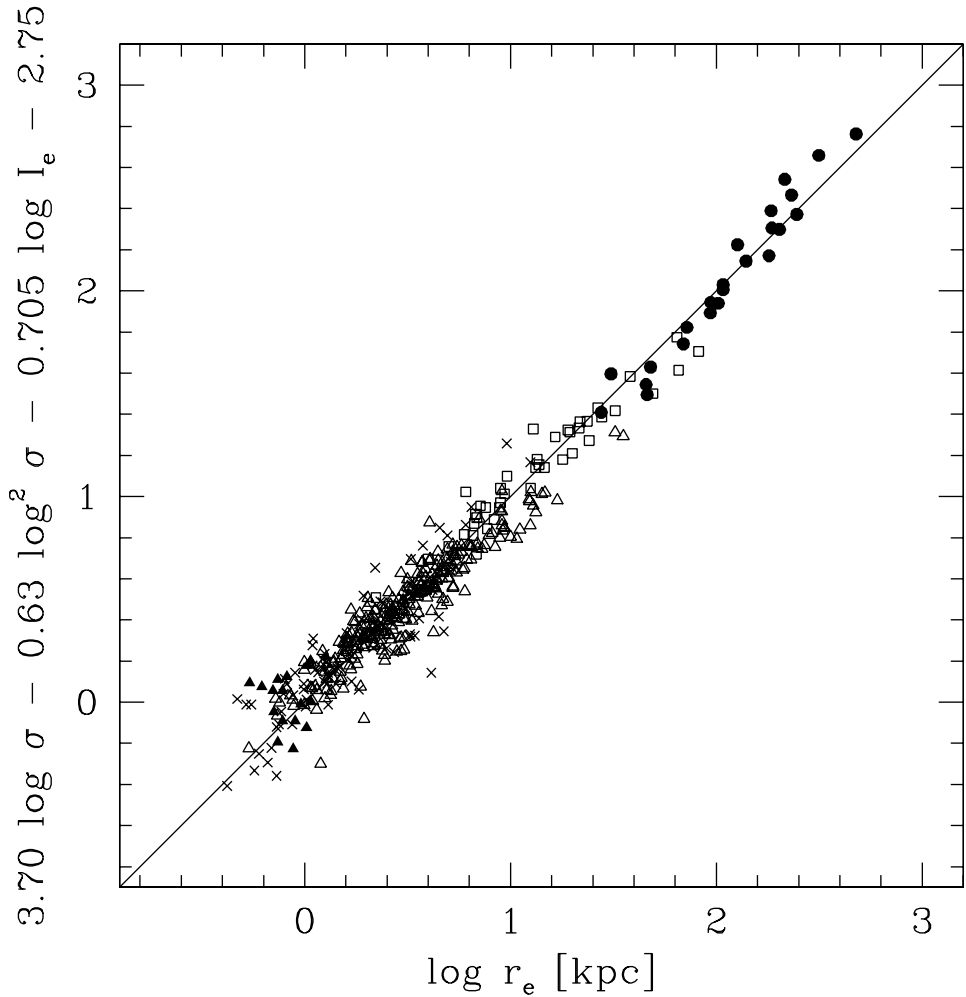}
\caption{Edge-on view of the 2-D manifold. 
Units given in text. All luminosities are corrected to the $I$-band
(see text). Solid circles represent our CSph measurements, open
squares the BCGs of \cite{oh}, open triangles the Es of \cite{jorg}, crosses the Es 
of \cite{coma}, and solid triangles the dEs of  \cite{geha}.  The line is the 1:1 line, and for
this choice of axes it represents the best fit manifold. 
}
\label{fig:fpall2}
\end{figure*}

We fit the data for dEs \citep{geha,coma}, Es \citep{jorg}, BCGs \citep{oh} and
CSphs to a slightly rearranged version of Eq. 7, 
which we term the ``fundamental manifold" in analogy to the
FP, 
$$\log r_e  = -\alpha^2 \log^2\sigma + (2+2\alpha\beta)\log\sigma + B\log I_e + C_2. \eqno(8)$$
The resulting parameters are $\alpha^2 = 0.63\pm0.02$, $(2+2\alpha\beta) = 3.70\pm 0.07$, 
$B = -0.705 \pm 0.004$ and $C^\prime = -2.75 \pm 0.08$. 
If one uses only the ICS luminosity  for the CSphs, then the resulting 
parameters are $\alpha^2 = 0.66\pm0.02$, $(2+2\alpha\beta) = 3.81\pm 0.08$, 
$B = -0.676 \pm 0.004$ and $C^\prime = -2.90 \pm 0.08$. 
The comparable plot to Figure \ref{fig:fpall1} is shown in Figure \ref{fig:fpall2} and the 
data used to obtain this fit are presented in Table \ref{tab:mdata}.  The
rms scatter along the $r_e$ axis is now 0.099 for the combination of the Es, BCGs and
CSphs (0.107 if the CSph is described by the ICS alone),  
which is not significantly worse than the 0.094 scatter of the elliptical FP 
alone. With the addition of the dwarf elliptical samples \citep{geha,coma}),
the scatter rises to 0.114.
The combined sample is highly weighted toward normal ellipticals, simply because
they dominate the sample numerically, so it is important to examine the scatter carefully.
For example, fitting a single FP to these same data results in a scatter of 0.129 between observed and fitted $r_e$, which 
is not much worse than that derived from the manifold fit. However, 
the rms deviation for the CSphs about the the manifold
is 0.095, while it is 0.207 for the single FP fit. At the 
other extreme in the $\sigma$ range, the dEs from the \cite{geha} sample have 
an rms deviation of 0.165 about the manifold and 
0.256 about the single FP. The manifold results in significant improvements at both
ends of the $\sigma$ range.\footnote{A comparison of the scatter of CSphs, as described
by either the ICS+BCG or ICS luminosity, about the 
fundamental manifold fits again favors the ICS+BCG description of the CSph
(the scatter in $\log r_e$ about the fundamental manifold fit, when allowing for a normalization error, is
0.095 for the ICS + BCG vs. 0.138 for the ICS, and 0.095 vs. 0.153, respectively, 
when not allowing for a normalization error).}

There are many reasons why our manifold fit must result in a scatter larger than
the intrinsic scatter:

First, we combine data from various surveys carried 
out in different filters with different analysis techniques. 
Although we attempt to correct the colors to the same
system, these corrections are
based on mean observed or modeled colors for the population and mean filter+detector 
transmission curves.

Second, we do not correct for any wavelength dependence
on $r_e$, which has been shown to exist at least for 
ellipticals \citep{pahre}. While this is a systematic error within a given sample, in the merger
of samples presented here it would increase the calculated scatter.

Third, we use the velocity dispersion of the
cluster galaxies rather than the ICS. Theoretical modeling \citep{murante, willman, sl} and observations \citep{zibetti}
conclude that the ICS is somewhat more centrally concentrated than the cluster galaxies, 
and hence could have somewhat different kinematics. \cite{sl}
suggest that at small radius 
the ICS could have about one-half the velocity dispersion of the galaxies. Indeed, a difference
between the velocity dispersion of the cluster galaxies and ICS has been detected in 
the Coma cluster \citep{gerhard}, although 
\cite{kelson} find $\sigma_{ICS} \rightarrow \sigma$ at large radii in another cluster. 
If $\sigma$ is roughly a constant fraction or multiple of $\sigma_{ICS}$,
then this difference would be absorbed into the constant term of Eq. 8. If
$\sigma$ is only crudely related to $\sigma_{ICS}$, then
we would find a large scatter in our relationships (making those relationships more difficult to identify). 
If the relationship between $\sigma$ and $\sigma_{ICS}$
depends on other cluster parameters, such
as the concentration, then this would introduce a tilt in our relationships 
and move the CSphs away from the BCG FP.
The continuity between the CSph and BCG FPs argues against this being a large effect.
In any case, we expect any problems in using the galaxy velocity dispersion as a proxy
for the ICS velocity dispersion to 
be an additional source of noise.

Fourth, 
we apply no correction for non-sphericity, orbital anisotropy 
\citep[cf.][]{ciotti},  or rotation
in any of these systems \citep[cf.][]{busarello}. 
While the rotation velocity of the ICS
is presumably small in comparison to the velocity dispersion, the large ellipticity of
the ICS components must at least necessitate a correction when converting from line-of-sight
velocity dispersion to the velocity dispersion applicable for the FP.  

Fifth, we do not
correct the measured velocity dispersion to the velocity dispersion
within $r_e$. As suggested by \cite{sl}, an additional complication is
that this correction may be different for the 
galaxies than for the ICS. 

Sixth, we ignore any direct dependence of $M_e/L_e$ on $I_e$ or $r_e$. 
To the degree that 
$\sigma$ can predict $I_e$ or $r_e$, we will model variations in $M_e/L_e$ with those parameters,
but we know from the fact that spheroids do not fall on a fundamental line that $\sigma$ alone
cannot reproduce the behavior in $I_e$ and $r_e$ exactly. In particular, we have not
accounted for $B \ne 1$ in our scenario. 

Seventh, we adopt the parabolic description of 
the relationship between $M_e/L_e$ and $\sigma$  as a mathematical convenience. The
correct functional form could be asymmetric, or have a flatter bottom than a parabola, or less
curvature. All of these issues suggest that the intrinsic scatter about the true ``fundamental manifold"
could be smaller than that measured here.

As with most explorations of the FP, one is left with intriguing relationships but no direct
understanding of the physics that gives rise to those relationships. At the core
of all of these fitting functions is the virial theorem, but it is the deviations from the simple expectations
that can illuminate patterns in galaxy formation. 
In our case, we have two distinct results to understand: (1) why 
does $M_e/L_e$ rise with increasing $\sigma$ for large $\sigma$ and (2) why does $M_e/L_e$
rise with decreasing $\sigma$ for small $\sigma$? The balance between 
these two, possibly distinct, physical causes will account for the existence of the minimum 
$M_e/L_e$ and its particular value.  
In broad terms, and in different contexts,  both of these trends have been noted before
(generally for global $M/L$),  and various
investigators have attempted to explain them \citep[cf.][]{babul,martin, benson, marinoni, vdb, dekel}). 
The behavior at low $\sigma$ has drawn more interest because 1)
the observational effects --- high $M_e/L_e$ \citep{mateo} and low effective
yields \citep{garnett, tremonti} --- have been known for a longer time and 2) the formation
history of low mass halos is critical in hierarchical models \citep{babul}. 
Two primary processes are cited as responsible for the comparatively low star
formation efficiency in low mass systems: winds  and UV photoionization.
One expects that supernova winds will more easily remove 
material in lower mass systems \citep{martin, dekel} and that 
the intergalactic UV radiation field will
more easily ionize the gas \citep{babul}. 

The behavior on the largest scales has been explored less, although \cite{pad} 
found an analogous rise in $M/L$ as a function of mass among Es, and various groups
have recently begun modeling the formation of the ICS \citep{murante, willman, sl}.
Those simulations predict that the ICS forms from the mergers of groups early
in the history of the cluster. Because the ICS baryons are already in the form of stars
at that stage, they behave much like the dark matter during the mergers. Therefore, they do not concentrate
toward the center of the potential well and do not form systems with low $M_e/L_e$.

We conclude that the trend between $M_e/L_e$ and $\sigma$ describes how well 
baryons are able to dissipate and concentrate toward the center of the 
potential. The various driving factors are the degrees to which 1) the baryons have
already been converted to stars prior to the assembly of the system, 2) new gas
can cool and form stars in the final system, and 3) winds expel gas
and lower the binding energy of the central component. Theoretical work is beginning to explore
how mergers affect the relative distribution of dark and luminous matter \citep{bk}.

\subsection{Interpreting the Minimum}

The choice of a parabola to describe the $\log M_e/L_e$ versus $\log \sigma$ relation 
implies a well-defined global minimum in the relationship; however, we have stressed
before that this particular functional form is somewhat arbitrary. Even so, the data,
particularly for the dSphs, do suggest
an upturn in $M_e/L_e$ at low $\sigma$, although a flat trough is not ruled out. 
If we accept the assumption of a parabolic relationship between $M_e/L_e$ and
$\sigma$, our best fit manifold from Eq. 7 implies $\alpha = 0.79\pm 0.03$ and
$\beta = 1.07 \pm 0.05$ and that the minimum $M_e/L_e$ occurs in systems
with  $\sigma =22.9 \pm 4.7$ km sec$^{-1}$.
The degree to which these values are given physical significance is in 
large part determined by the validity of the mathematical form we have chosen,
how well the various samples have been placed on the same photometric
and kinematic systems, and the sampling of the data through the region of the minimum.
Unfortunately, the bulk of our data are at larger $\sigma$ than the minimum
and so our fit is quite sensitive to the limited data available near the minimum.

In contrast to what we did above, the turnover in the
$M_e/L_e$ resulting from dimensional arguments (\S 3.4) occurs at $\sigma \sim 62$ km 
sec$^{-1}$ (Figure \ref{fig:mtol2}). While we have argued that the
evaluation of $M_e/L_e$ using homology and dimensional arguments 
is not internally self-consistent with our modeling
of the fundamental manifold (because of the higher order dependence of $\log M_e/L_e$
on $\log \sigma$ that we impose), these estimated values of $M_e/L_e$ do reflect something
about the relationship between $\sigma$, 
$r_e$, and $I_e$. The reason the estimates do not agree in detail is presumably because the correct
coefficients are not available in the dimensional argument and because the homology assumption
breaks down.
Which turnover is the more relevant in assessing the relative effects of winds,
or photoionization, or structural changes due to the degree of dissipation is 
unclear. We conclude that while a change in behavior does occur, the
exact value of $\sigma$ ranges between at least 20 to 60 km sec$^{-1}$. The corresponding
mass scale is much lower than that suggested by studies of the global $M/L$ that
also find a minimum $M/L$ at a specific mass scale \citep{benson, marinoni, vdb}, but
the results are not necessarily in conflict.
Observing differences in the behavior of $M/L$ within different radii, in this case 
within $r_e$ versus $r_{200}$,
could potentially be used to break the degeneracy in the analysis between the dissipation of
baryons and the efficiency of converting them to luminous matter.

\subsection{Caveats}

As with all analyses of the FP and its cousins, the description of 
spheroids as fully virialized, homologous (or at least weakly homologous) dynamical systems
is oversimplistic. Our hope is that a minimalist description
will embody the principal elements of spheroids, and the mere existence of the FP 
supports this approach. Nevertheless, to ignore the richness of the process of 
spheroid formation could lead to an overinterpretation of fortuitous 
coincidences. Even in the case of low scatter in the FP ($\sim$0.1), the 
scatter between observed and model $r_e$ is still 26\%, 
which, while small considering the range of $r_e$ values, is not negligible.
For the analysis and discussion we have presented, we remain concerned about the following:

\medskip
\noindent
We have ignored possible variations of $M_e/L_e$ that depend on $r_e$ and $I_e$.
The standard approach attributes the tilt in the elliptical FP
 as being partly due to $\sigma$ and partly to the other terms in Eq. 4.
We justify placing the entire burden on the form of the $\sigma$ term on
the basis
of the large variation in $A$ among FPs and the success of the model. However, given the
various correlations between $r_e$, $I_e$, and $\sigma$ it is possible to slough off some
of that variation to the other parameters. In a complete treatment, one would aim
to explain the behavior of the $B$ coefficient in Eq. 8 as well as that of $A$. However,
examining the residuals from 
the best fit manifold versus $I_e$, we find only a marginal indication of systematic behavior, and hence
expect a minimal reduction in the scatter with the addition of a $I_e$-dependent term.
Instead, we choose to limit the number of new parameters introduced at this stage, but, as samples increase,
particularly for extremely low and high $\sigma$ spheroids,
we anticipate more complex and complete modeling.

\medskip
\noindent
We have ignored other sources of homology breaking in our analysis. For example,
systematic changes in orbital structure or in the dark matter potential as a function of 
spheroid scale will also tilt the FP. We argue that such changes cannot
account for the large (factor of $\sim$ 100) difference in our calculated $M_e/L_e$ across the entire
range of spheroids, but they are presumably there at some level. To the degree that
they vary as a low-order function of $\sigma$, they are incorporated into Eq. 8.

\begin{figure}
\plotone{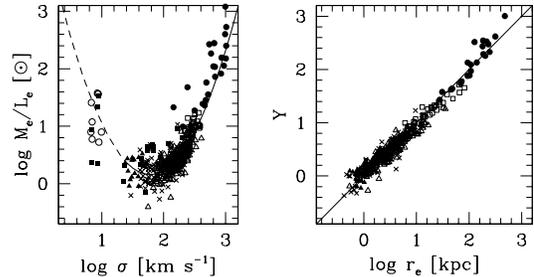}
\caption{Same as Figures \ref{fig:mtol2} and \ref{fig:fpall2} except the CSph luminosity
is taken to be only that of the ICS, rather than that of the ICS+BCG. In the left panel, the
solid line is the best fit parabola to the data shown in Figure \ref{fig:mtol2} and is included
here as a reference. In the right panel, $Y = 3.81 \log^2 \sigma - 0.66 \log \sigma - 0.68 \log I_e -2.90$,
and represents the best fit surface. Symbols are as described in Figure \ref{fig:mtol2}.
Whether the CSph is described by the ICS or ICS+BCG does not affect either the 
relationship between $M_e/L_e$ and $\sigma$ or the nature of the fundamental manifold.}
\label{fig:icsonly}
\end{figure}

\medskip
\noindent
We have focused on using the combination of BCG + ICS luminosity to describe the
CSphs. In no other type of spheroid did we explicitly combine two components. Are CSphs truly
described best by the combination, or are we being misled by small number statistics? Or
do other systems, such as normal ellipticals, actually have a low surface
brightness outer component, analogous to the ICS, that has not been observed?
Regardless of whether we study the ICS alone, or in combination with the BCG, the
existence of a fundamental manifold and the systematic behavior of $M_e/L_e$ with $\sigma$
remain (Figure \ref{fig:icsonly}).

\medskip
\noindent
The trends in $A$ and in $M_e/L_e$ are dominated by inter-sample differences rather
than by intra-sample differences. Because the samples come from disparate analyses,
we are concerned that some of the difference among them is due to systematic
errors rather than physical causes. Because there is significant overlap between
samples, we conclude that this is not a serious problem, but it is a fact that each of the principal samples alone is accurately described
by a FP relation. Marginal evidence for curvature was noted by \cite{jorg} in their sample, 
although \cite{bernardi} accounted for all the apparent curvature in their sample through
selection effects.

\medskip
\noindent
Because most of the available data is on the high $\sigma$ side of the $M_e/L_e$ minimum,
we have little constraint on the functional form of the $M_e/L_e-\sigma$ relation.
For example, an equally reasonable attempt at unification could come by modeling
the $M_e/L_e$ variation as a function of enclosed mass, $r_e\sigma^2$. This
approach leads to an equation that is somewhat harder to interpret because of cross-terms,
but that has the same number of free parameters. Even if the functional form, whatever it
may be,
is steeply rising on both sides (as reinforced by the dSph data), there is
no reason to believe that the rise should be symmetric. Once one allows for asymmetry,
the position of the minimum becomes quite uncertain and the need for data at
$\log \sigma < 1.5$ acute. 

\medskip
\noindent
The interpretation of $r_e$ depends on the surface brightness profile. There is still
some uncertainty in the literature regarding the relative merits of Sers\'ic versus deVaucouleur profiles
across all spheroid classes.
The latter type of profile is a special case of the former, so the more general statement that Sersi\'c
profiles fit spheroids is less controversial. A concern here is that there is a systematic
trend in profile shape with $\sigma$ that produces corresponding
systematic errors in $r_e$ and $I_e$, and hence in the FP.  \cite{bertin} have explored
this topic and concluded that  ``weak" homology is satisfied, but the exact
effect on an analysis such as ours is unclear. \cite{gg} find a continuous relationship
in structural parameters of dEs to Es when they fit Sers\'ic profiles rather than $r^{1/4}$ profiles.
Which profile is more appropriate across the full range of spheroids
and how to compare among objects, different studies, and different methods are open 
questions. However, profile differences seem unlikely to result in the large FP differences
observed among the full range of spheroids.

\section{Conclusions}
\label{sec:conclusions}

We demonstrate that spheroids ranging from dEs to the intracluster stellar component of galaxy clusters,
which we designate the cluster spheroid (CSph), 
lie on a curved surface, a 2-D manifold, in $(\sigma, r_e, I_e)-$space. Previous studies have
generally examined limited regions of parameter space, over which the manifold can be
adequately described by a plane.  Our principal findings are:

\noindent
$\bullet$ The addition of cluster velocity dispersions, as measured from the cluster galaxies,
decreases the scatter of the $r_e$-$I_e$ relation for the cluster spheroids, and therefore signifies the
existence of a ``fundamental plane" for clusters.

\noindent
$\bullet$ Combining the luminosity of the brightest cluster galaxy (BCG) and the intracluster stars (ICS)
leads to significantly lower
scatter in the CSph FP relationship. We argue that this effect is real, implying a connection between
the evolution of the BCG and ICS, rather than a
poor decomposition of the two components. Our results do not change qualitatively
if we use only the ICS for the CSph, but the scatter in various relationships increases
significantly. Unless otherwise noted, we refer to the combination of BCG and ICS as
the CSph, but in the more important cases present results for both descriptions of the CSph.

\noindent
$\bullet$ The scatter of the CSph FP (0.074) is as small as 
that observed for elliptical galaxies, but the orientation of the plane is different.

\noindent
$\bullet$ There is a systematic decline in the coefficient of the $\log \sigma$ term in
the equation of the FP as one progresses from systems with smaller to larger $\sigma$.
This trend suggests that a 2-D manifold, which over the limited range of $\sigma$ 
probed by any one spheroid population is indistinguishable
from a plane, roughly forms a set of twisted planes that  fit spheroids from 
low mass ellipticals to the CSphs of the most massive clusters.

\noindent
$\bullet$
The non-linear relation between $\log M_e/L_e$ and $\sigma$ for spheroids implies a functional
relation between these quantities that is more complex than a power law,  and so requires
the breaking of the homology assumption. 
We adopt $\log M_e/L_e = (\alpha \log \sigma - \beta)^2 + \gamma$ as a model of this
relationship because it reproduces
the systematic decline in the coefficient of the 
$\log \sigma$ term in the FP equation and captures the
curvature seen  in the $\log M_e/L_e$ versus $\log \sigma$ relationship for spheroids.
Of particular importance is that 
$M_e/L_e$ represents the mass to light ratio within $r_e$ and may be decoupled
from the global $M/L$. The magnitude
of the change in $M_e/L_e$ across the full range of spheroids
excludes changes in stellar populations as a possible explanation for the global trend. 
Instead, these changes must be driven primarily by variations in the relative amounts of luminous
and dark matter within $r_e$, although some modest dependence of $M_e/L_e$ with stellar populations
must exist (see \cite{cap} for a recent exploration of this topic for E's).

\noindent
$\bullet$
The resulting 2-D manifold has rms residuals along the $r_e$
axis of 0.099 (for Es, BCGs, and CSphs),  compared to 0.094 for the FP plane fit to ellipticals alone and 
to 0.089 for BCGs alone. If we extend the manifold to dEs, the scatter increases slightly (0.114).
The increased scatter relative to the individual FP's of each spheroid population
is insignificant given that the manifold fit is done with 
respect to five different samples that were observed in different photometric bands by different investigators. Using the best fit 2-D manifold to define the coefficients of the relationship between 
$\log M_e/L_e$
and $\log \sigma$, we find a minimum $M_e/L_e$ for galaxies with $\sigma = 22.9 \pm 4.7$
km s$^{-1}$. We qualitatively agree with a range of theoretical and
observational studies that conclude that there exists a galaxy mass scale for which $M/L$ is minimized
\citep{benson, marinoni, vdb}, although 
our value for that mass scale is significantly
lower. Part of the discrepancy may arise from the
measurement of this phenomenon on global scales, $r_{200}$, 
versus within $r_e$.  

Our new description that places the family of spheroids on a 2-D manifold in  
$(\sigma, r_e, I_e)-$space is remarkable in that it unifies spheroids that span a factor of 100 in $\sigma$ and 
1000 in $r_e$. This range of systems must have different histories both in terms of
when they were assembled and when their stars formed.
Even so, the global properties of these systems end up on 
the manifold with intrinsic scatter that must be less than 0.114 (30\% in $r_e$). 
We are able to place all dark matter dominated spheroids on a single
relation by accounting for the observed non-linear relationship 
between $\log(M_e/L_e)$ and $\log \sigma$. This relationship,
which deviates from simple virial theorem and homology expectations, points to a 
more  specific phrasing of the question of how galaxies and clusters form:
what is the origin of the relationship between $M_e/L_e$ and potential well depth?
The relative importance of dissipation, cooling, star formation,
feedback, and dynamical relaxation during spheroid formation 
must vary with spheroid scale in such a manner as to 
reproduce the continuous trend observed in $M_e/L_e$ (Figure \ref{fig:mtol2}), a 
trend responsible for the ``fundamental manifold" (Figure \ref{fig:fpall2}).
Theoretical explorations, such as those focusing on 
the interplay between different feedback mechanisms and the growth of galaxies  
\citep[for examples, see][]{pad, bk, dekel2}, 
may ultimately reveal the nature of the observed trend in 
$M_e/L_e$ and thereby unify the formation processes of spheroids.
  
\begin{acknowledgments}
The authors thank I. J\o rgensen  and A. Matkovic for providing their data for this study, and D. Eisenstein
for a discussion of homology. 
DZ acknowledges financial support for this work from the David and Lucile Packard Foundation,
NASA LTSA award NNG05GE82G and NSF grant AST-0307482.
AHG is funded by an NSF Astronomy \& Astrophysics Postdoctoral Fellowship
under award AST-0407085. AIZ acknowledges financial support from NASA LTSA award NAG5-11108 and
from NSF grant AST-0206084.
This research has made extensive use of  NASA's Astrophysics Data
System and the NASA/IPAC Extragalactic Database (NED), which is operated by
the Jet Propulsion Laboratory, California Institute of Technology, under contract
with NASA. 
\end{acknowledgments}

\clearpage
\end{document}